\documentclass[%
 reprint,
superscriptaddress,
 amsmath,amssymb,
 aps,
]{revtex4-2}

\usepackage{xcolor}
\usepackage{graphicx}
\usepackage{dcolumn}
\usepackage{bm}

\begin{document}

\preprint{APS/123-QED}

\title{Seeing new depths: Three-dimensional flow of a free-swimming alga} 


\author{Gregorius Pradipta}
\affiliation{Department of Chemical Engineering and Materials Science, University of Minnesota, Minneapolis, MN 55455, USA}

\author{Wanho Lee}
\affiliation{National Institute for Mathematical Sciences, Daejeon 34047, Republic of Korea}%

\author{Van Tran}
\affiliation{Department of Chemical Engineering and Materials Science, University of Minnesota, Minneapolis, MN 55455, USA}%

\author{Kyle Welch}
\affiliation{Department of Chemical Engineering and Materials Science, University of Minnesota, Minneapolis, MN 55455, USA}

\author{Santosh K. Sankar}
\affiliation{Department of Mechanical Engineering, University of Minnesota, Minneapolis, MN 55455, USA}

\author{Yongsam Kim}
\affiliation{Department of Mathematics, Chung-Ang University, Seoul 06974, Republic of Korea}

\author{Satish Kumar}
\affiliation{Department of Chemical Engineering and Materials Science, University of Minnesota, Minneapolis, MN 55455, USA}

\author{Xin Yong}
\affiliation{Department of Mechanical and Aerospace Engineering, University at Buffalo, Buffalo NY 14260, USA}

\author{Jiarong Hong}
\affiliation{Department of Mechanical Engineering, University of Minnesota, Minneapolis, MN 55455, USA}
\affiliation{Saint Anthony Falls Laboratory, University of Minnesota, Minneapolis, MN 55414, USA}

\author{Sookkyung Lim}
\affiliation{Department of Mathematical Sciences, University of Cincinnati, Cincinnati, OH 45221, USA}

\author{Xiang Cheng}
\affiliation{Department of Chemical Engineering and Materials Science, University of Minnesota, Minneapolis, MN 55455, USA}
\affiliation{Saint Anthony Falls Laboratory, University of Minnesota, Minneapolis, MN 55414, USA}


\date{\today}

\begin{abstract}
A swimming microorganism stirs the surrounding fluid, creating a flow field that governs not only its locomotion and nutrient uptake, but also its interactions with other microorganisms and the environment. Despite its fundamental importance, capturing this flow field and unraveling its biological implications remains a challenge. Here, we report the first direct, time-resolved measurements of the three-dimensional (3D) flow field generated by a single, free-swimming microalga, \textit{Chlamydomonas reinhardtii}, a model organism for microbial locomotion and flagellar dynamics. Supported by hydrodynamic modeling and simulations, our measurements resolve how established two-dimensional (2D) flow features such as in-plane vortices and the stagnation point emerge from and shape the full algal flow in 3D. Moreover, we reveal unexpected low-Reynolds-number flow phenomena including micron-sized vortex rings and periodically recurring translating vortices and uncover topological changes in the underlying flow structure associated with the puller-to-pusher transition of an alga. Biologically, access to the 3D flow field enables rigorous quantification of the alga's energy expenditure, as well as its swimming and feeding efficiency, improving the precision of these physiological metrics. Taken together, our study demonstrates rich vortex dynamics in inertialess flows and shows their influence on microbial motility. The work also introduces a new experimental method for mapping the fluid environment sculpted by beating flagella.
\end{abstract}

\maketitle


\section{Introduction}
To swim, a microorganism must move fluid around itself, creating a flow field with complex spatiotemporal patterns that extend over a region much larger than the size of the microorganism \cite{Dusenbery_2009_book,Lauga_2020_book}. Far from being merely a byproduct of locomotion, this flow field governs a wide range of microbiological processes, such as nutrient uptake \cite{Magar_2003_nutrient,Tam_2011_optimization}, the detection of and communication with other microorganisms \cite{Sheng_2007_detection,Woolley_2009_communication}, the perception and adaptation to changing environments \cite{Berke_2008_surface,Wheeler_2019_flow,Buchner_2021_surface}, the rheology of microorganism suspensions \cite{Saintillan_2018_rheology}, and the emergence of collective multi-cellular structures \cite{Durham_2009_algae,Peng_2021_turbulence,Liu_2021_turbulence,Shekhar_2025_colonial}. Hence, the flow field of a swimming microorganism is regarded as one of its fundamental characteristics \cite{Lauga_2009_review,Marchetti_2013_review}. More broadly, resolving the flow field around swimming microorganisms would also shed light on the basic function of flagella and cilia in fluid transport, essential for biological processes across all three domains of life \cite{Khan_2018_fagella,Gilpin_2020_NatureReview}. 

Despite its significance, the flow field of free-swimming microorganisms remains poorly understood due to the challenge of imaging three-dimensional (3D) fluid flow at micron scales with millisecond precision around a fast, irregularly moving object. Establishing a milestone in experimental biofluid mechanics \cite{Saintillan_2010_comment}, pioneering studies using bright-field microscopy have captured two-dimensional (2D) projections of the 3D flow field around a freely swimming alga, \textit{Chlamydomonas reinhardtii} (\textit{C. reinhardtii}), revealing intricate flow patterns that have greatly enhanced our understanding of microbial motility \cite{Drescher_2010_algalflow,Guasto_2010_algalflow}. Similar techniques have later been applied to obtain the 2D flow around freely swimming bacteria \cite{Drescher_2011_Ecoliflow}. Simulations \cite{Fauci_1993_algae,OMalley_2012_simulation,Klindt_2015_simulation,Li_2017_simulation} and more recent experiments \cite{Jeanneret_2019_confinement,Mondal_2021_confinement} on \textit{C. reinhardtii} have also largely focused on its 2D flow features. A systematic investigation of the 3D flow of this model microorganism remains lacking. Indeed, experimental measurements of the 3D flow field around \textit{any} freely swimming unicellular microorganism have not yet been achieved to date. 

As the swimming gaits of many microorganisms including \textit{C. reinhardtii} lack axial symmetry, the 2D flow fields fail to capture the full complexity of the 3D flow structures and their profound implications for microbial physiology. For example, the time-averaged 2D flow field of \textit{C. reinhardtii} shows two vortices flanking the cell body and a stagnation point in front [see e.g. Fig.~4 in \cite{Drescher_2010_algalflow} and Fig.~2 in \cite{Guasto_2010_algalflow}]---features now recognized as hallmarks of algal swimming. While intriguing, these 2D features raise questions about the underlying nature of the algal flow: What is the relationship between the two lateral vortices---are they independent structures, or are they connected in 3D? What role does the stagnation point play in redirecting flow outside the imaging plane? Time-resolved measurements further reveal the migration of vortices alongside the alga, in coordination with periodic switching between puller and pusher swimming modes \cite{Guasto_2010_algalflow,Klindt_2015_simulation}. Does the puller-to-pusher transition trigger a topological change in the underlying flow structure \cite{Ricca_1996_flowtopology,Yao_2022_vortexringreview}? If so, how does such a change unfold?  Two-dimensional flow fields alone are insufficient to address these questions, which demand a detailed examination of the 3D flow field around \textit{C. reinhardtii}. Beyond its intrinsic relevance to microbial fluid mechanics, the 3D flow is also critical for evaluating key biological functions of algae, such as energy expenditure, swimming efficiency, and nutrient uptake, which have so far been analyzed primarily using 2D flow data \cite{Guasto_2010_algalflow,Tam_2011_optimization}.

Here, we present a direct experimental measurement of the 3D flow field of a freely swimming \textit{C. reinhardtii}. Combining experiments with hydrodynamic modeling and simulations, we fill the knowledge gap at the intersection of fluid mechanics and microbiology. Our study uncovers unexpectedly rich structures in the flow of this model organism, expanding the understanding of both low-Reynolds-number hydrodynamics and their biological consequences for microbial motility and physiology.

\section{Experiment}

\begin{figure}[t]
    \includegraphics[width=1\linewidth]{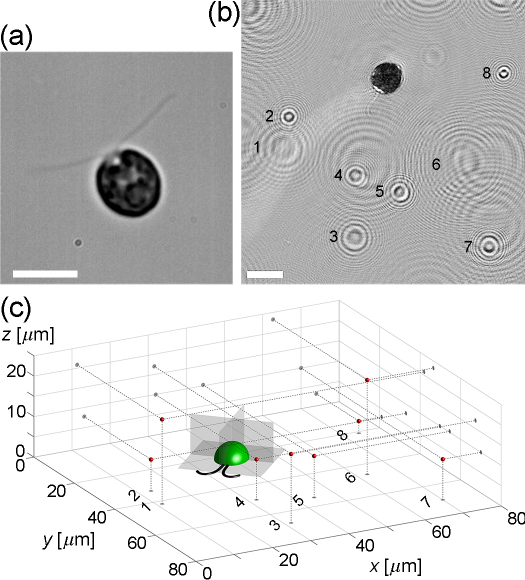}
    \caption{Imaging \textit{C. reinhardtii}. (a) A bright-field micrograph of a unicellular alga, \textit{C. reinhardtii}. (b) A hologram of a freely swimming alga in a dilute suspension of 1-$\mu$m-diameter polystyrene tracers. Scale bars: 10 $\mu$m. (c) Reconstruction of the hologram revealing the 3D positions of the tracers. Eight tracers labeled by number are identified within 25 $\mu$m of the focal plane. The three orthogonal planes through the center of the alga, shown in Fig.~\ref{fig:figure2}, are indicated.}
    \label{fig:figure1}
\end{figure}

\begin{figure*}[t]
    \centering
    \includegraphics[width=0.81\linewidth]{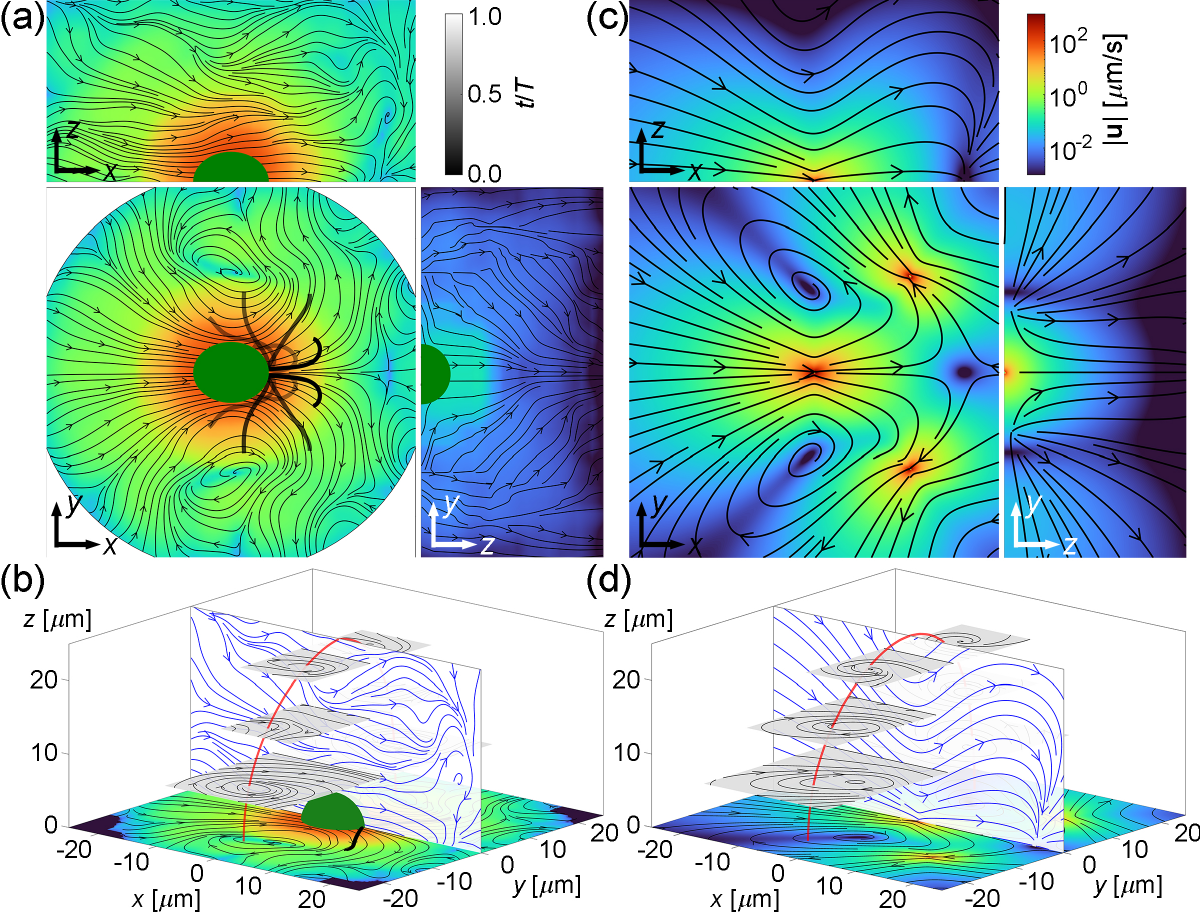}
    \caption{Time-averaged 3D flow field from experiments (a,b) and the three-Stokeslet model (c,d). (a) Flow fields in the $x$–$y$, $x$–$z$, and $y$–$z$ planes passing through the center of the algal body (dark green ellipsoid). The semi-major and semi-minor axes of the body are 5 $\mu$m and 4 $\mu$m, respectively. Streamlines (black lines) indicate flow direction, and color represents the magnitude of in-plane velocity. Flagellar strokes over one beat cycle, extracted from experiments, are overlaid in the $x$–$y$ plane, with phase indicated by grayscale shading; $t = 0$ marks the start of the power stroke. (b) 3D rendering of the experimental flow field. The red line traces a vortex line through the core of the lateral vortices. (c,d) Corresponding flow fields from a model of three Stokeslets between two parallel walls located 32 $\mu$m above and below the alga in the $z$-direction (SM Sec.~1A).}
    \label{fig:figure2}
\end{figure*}

We use high-speed digital in-line holographic microscopy to image the time-resolved 3D flow induced by a single freely swimming unicellular alga, \textit{C. reinhardtii}, which is chosen because of its wide use as a model for eukaryotic motility and flagellar dynamics \cite{Guasto_2012_review,Goldstein_2015_review,Jeanneret_2016_review,Leptos_2023_algaephototaxis} and its well-established 2D flow field for comparison \cite{Drescher_2010_algalflow,Guasto_2010_algalflow}. A cell of \textit{C. reinhardtii} has an approximately prolate spheroidal body with a semi-minor axis $a \approx 4$ $\mu$m and a semi-major axis $c \approx 5$ $\mu$m (Fig.~\ref{fig:figure1}(a), Appendix A). The cell swims by beating two anterior flagella of length $l \approx 12$ $\mu$m in a breaststroke-like manner at a frequency $f \approx 50$ Hz or a period $T=1/f \approx 20$ ms (Fig.~\ref{fig:figure2}(a)). The beating flagella propel the cell in an oscillatory manner with a mean speed $\langle U \rangle \approx 115$ $\mu$m/s. The Reynolds number of the flow of a swimming \textit{C. reinhardtii} is $Re= 2 \rho \langle U \rangle c /\eta \approx 1.15 \times 10^{-3}$, where $\rho = 10^3$ kg/m$^3$ and $\eta  = 1$ mPa$\cdot$s are the density and viscosity of water, respectively.

We add a low volume fraction (0.02\% v/v) of 1-$\mu$m diameter polystyrene (PS) spheres as tracers in a dilute algal suspension ($\sim 2000$ cells/ml). A collimated laser beam of 452 nm is used for illumination. The light scattered by PS tracers interacts with the incoming light, which leads to interference patterns---a hologram—containing the information of 3D tracer positions (Fig.~\ref{fig:figure1}(b)) \cite{Kumar_2025_Holographyreview}. While the hologram is taken at 500 frames per second, its 3D reconstruction is achieved numerically offline (Fig.~\ref{fig:figure1}(c)) \cite{Toloui_2015_holography}. A standard particle-tracking algorithm is then applied to extract the 3D tracer trajectories around a swimming alga. We analyze tracer motions only when an alga swims with its two flagella beating symmetrically within the focal plane, several tens of microns away from the system boundaries. Details of our experimental protocol can be found in Appendices B and C. 

\section{3D Flow Field}

We present both the time-averaged and time-resolved 3D flow fields of a freely swimming alga. The physiological implications of these flow fields are then discussed in the next section.

\subsection{Time-averaged flow field} 

Figures~\ref{fig:figure2}(a) and (b) show the time-averaged 3D flow field of a single alga in the laboratory frame. The Cartesian coordinate is defined with the alga swimming in the $+x$ direction, its body centered at the origin, and its two flagella beating in the $x$-$y$ plane. The flow in the $x$-$y$ plane shows a hyperbolic stagnation point in front of the alga and two lateral vortices, closely matching the reported 2D-projected flow fields \cite{Drescher_2010_algalflow,Guasto_2010_algalflow}. The 3D visualization enables us to explore the flow features in the $x$-$z$ and $y$-$z$ planes that were not captured in prior work. In the $x$-$z$ plane, the flow tilts slightly toward the algal body behind the alga and exhibits a hairpin-like structure in front, where it diverges from the flagellar plane, then loops back and converges at the stagnation point. Since the alga moves along the $x$ direction, the flow in the $y$-$z$ plane is substantially weaker. Interestingly, two fluid sources are identified on either side of the alga, ejecting fluid out of the flagellar plane.

It has been shown that the 2D time-averaged flow can be described by the flow field of three Stokeslets, with one Stokeslet at the cell body pushing forward and one each near the flagella pushing backward \cite{Drescher_2010_algalflow}. The summation of the three forces is zero, ensuring the force-free condition of the free-swimming alga. We extend the model to consider three Stokeslets under weak confinement between two parallel walls \cite{Liron_1976_confinedStokeslet}, accounting for the influence of system boundaries in our experiments (Supplementary Material (SM), Sec.~1A) \cite{supplemental}. The simple model qualitatively captures the 3D structure of the time-averaged flow, showing the hairpin-like flow in the $x$-$z$ plane and fluid sources in the $y$-$z$ plane (Fig.~\ref{fig:figure2}(c)).

\begin{figure*}[t]
    \centering
    \includegraphics[width=1\linewidth]{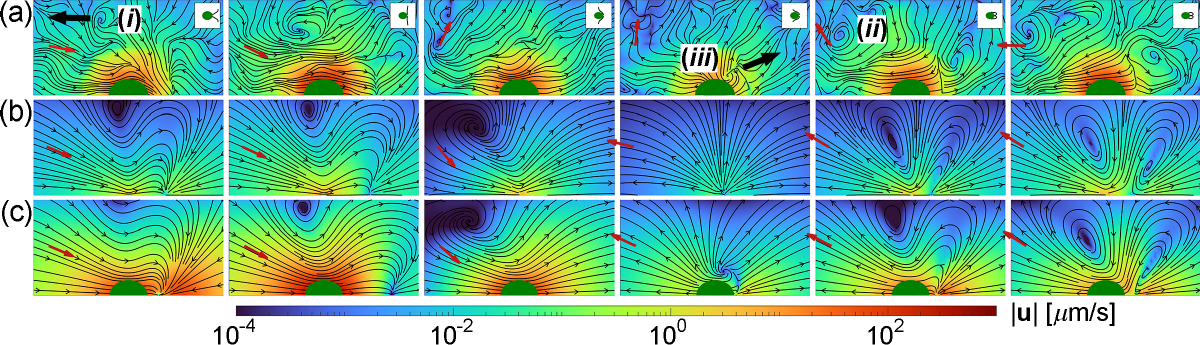}
    \caption{Time-resolved 3D flow fields from experiments (a), the dynamic three-Stokeslet model (b), and regularized Stokeslets simulations (c). Flow fields in the $x$–$z$ plane through the center of the alga are shown at different beat phases, as indicated by the flagellar shapes (upper-right insets in (a)). Red arrows denote local flow velocities, which exhibit counterclockwise (CCW) rotation. Three vortices---labeled Vortex (i), (ii), and (iii)---are marked at the phases when they first emerge in panel (a), with their propagation directions indicated by thick black arrows. All flow fields are computed with confinement from parallel walls located 32 $\mu$m above and below the alga. The semi-major axis of the algal body (dark green semi-ellipsoid) is 5 $\mu$m. See also Movie S1.}
    \label{fig:figure3}
\end{figure*}

The 3D flow fields revealed by our experiments and hydrodynamic modeling resolve how previously identified 2D features govern the full structure of the algal flow. In particular, the stagnation point in the $x$–$y$ plane draws fluid inward in the $z$ direction, producing a local uniaxial extensional flow along $y$, qualitatively different from a planar extensional flow one might infer from the 2D field alone. This 3D structure of the stagnation point would influence the entrainment of passive particles during their close encounters with the alga \cite{Jeanneret_2016_entrainment}. More interestingly, the lateral vortices connect outside the flagellar plane to form a closed vortex ring with its central axis aligned along $+x$ within the flagellar plane. The core of the ring is highlighted by the red line in Fig.~\ref{fig:figure2}(b) and (d) (Appendix D). Vortex rings are typically associated with intermediate to high $Re$ at large scales \cite{Sullivan2008_vortexring,shariff1992vortex}. With a ring diameter of $\sim 12$ $\mu$m, our measurements reveal the smallest known vortex ring at vanishing $Re$. While Fig.~\ref{fig:figure2}(b) represents a vortex ring in the time-averaged flow, micro-vortex rings and their dynamics are captured by the time-resolved flow field, as shown next.

\subsection{Time-resolved flow field}

The time-resolved 3D flow field from holography reveals even more intriguing spatiotemporal patterns.

\subsubsection{Experiment} 
Figure~\ref{fig:figure3}(a) illustrates the temporal evolution of the flow in the $x$-$z$ plane through the center of the alga, which reveals two notable interdependent features. First, behind the algal body, the flow exhibits counterclockwise (CCW) rotation at the flagellar beating frequency $f$ (red arrows in Fig.~\ref{fig:figure3}(a)). The rotation speed is non-uniform, with the fastest rotation occurring during the transition from the power to the recovery stroke (SM Fig.~S1) \cite{supplemental}. Second, we observe three distinct vortices within one beat cycle. Two vortices form behind the body at approximately 10 to 11 o'clock: Vortex (i) emerges during the power stroke, rotating CCW and traveling downstream during the power stroke, whereas Vortex (ii) appears during the recovery stroke, rotating clockwise (CW) with less obvious movement. The formation and propagation of these rear vortices drive the observed CCW rotation of the flow. A third vortex [Vortex (iii)] forms in front of the body, rotating CCW and moving forward during the recovery stroke.

\subsubsection{Modeling} 

The traveling vortices trailing behind the algal body is reminiscent of vortex shedding at high $Re$, a fluid phenomenon central to the locomotion of macroscopic organisms such as insects, birds, and fish \cite{Wu_2011_vortexshedding,Liu_2024_vortexshedding}. This ``vortex shedding'' by the alga cannot be explained by a simple extension of the three-Stokeslet model. When the magnitudes of the three Stokeslets are oscillated at frequency $f$ with force determined by resistive force theory (RFT) from the experimentally extracted shape of the flagella (SM Sec.~1A and Fig.~S2) \cite{supplemental}, the flow velocities periodically vary in magnitude and sign (Movie S1). However, apart from direction reversal, the streamlines of the flow field remain stationary without either flow rotation or moving vortices. 

The observed vortex dynamics arise from the beating of the flagella. To mimic this process, we construct a dynamic three-Stokeslet model in which two anterior Stokeslets, representing the flagella, move along closed orbits determined by the force centroid of the flagella (SM Sec.~1B, Fig.~S2) \cite{supplemental,Brumley2014synchronization}. The flow field of the dynamic model displays qualitatively similar vortex dynamics as observed in experiments including the non-uniform flow rotation (SM Fig.~S1) \cite{supplemental}, the backward traveling CCW Vortex (i), nearly stationary CW Vortex (ii), and the forward moving CCW Vortex (iii) (Fig.~\ref{fig:figure3}(b), Movie S1). 

To more accurately capture the experimental flow field, we also employ the method of regularized Stokeslets \cite{Cortez_2001_RS,Cortez_2005_RS,Lee_2021_RSbacteria} to simulate the algal flow in the zero-$Re$-number limit (Fig.~\ref{fig:figure3}(c)), incorporating the shape of the algal body and the kinematics of the flagella extracted from experiments (SM Sec.~2A) \cite{supplemental}. The simulations predict an average swimming speed of $\langle U\rangle = 113$ $\mu$m/s, closely matching the experimental value. The time-averaged flow from the simulations also agrees well with the experimental measurements (SM Fig.~S6) \cite{supplemental}. The simulations reproduce all dynamic features of the experimental flow (Fig.~\ref{fig:figure3}(c), Movie S1). The quantitative differences likely stem from the substantial noise in the experimental time-resolved flow field (Appendices B and C). Hence, while experiments provide a benchmark for validating the simulations, the smooth flow fields from the simulations allow the calculation of higher-order velocity derivatives, enabling quantitative analysis of flow vorticity.

Before delving into flow vorticity and 3D vortex dynamics, it is worth commenting on the role of fluid inertia in shaping the algal flow. Fluid inertia is essential for vortex shedding in high-$Re$-number locomotion \cite{Wu_2011_vortexshedding,Liu_2024_vortexshedding}. Recent simulations and experiments on \textit{C. reinhardtii} have also identified the pronounced inertial effect in the flow of swimming algae \cite{Klindt_2015_simulation,Wei_2019_inertia,Wei_2021_inertia}. In particular, a fundamental singular solution of the unsteady Stokes equations---the Oscillet---has been proposed to model the vortex dynamics in the flow around a micropipette-held alga \cite{Wei_2021_inertia}. However, we find that the flow predicted by a model of three Oscillets fails to capture the key features of our experiments (SM Sec.~1C, Movie S1) \cite{supplemental}. Even more convincingly, immersed boundary simulations of the full Navier–Stokes equations \cite{Peskin_1996_IB} yield results that are indistinguishable from those obtained from simulations of regularized Stokeslets (SM Sec.~2A) \cite{supplemental}, confirming that fluid inertia has no significant effect on the vortex dynamics observed in the near-field flow of the alga. Thus, the periodically recurring translating vortices observed in the algal flow are fundamentally different from those associated with vortex shedding or a von K\'arm\'an vortex street at high $Re$ \cite{Guyon2015_hydrodynamics}. The algal vortices depend solely on the instantaneous configuration and velocity of the flagellum. When the flagella stop beating, they decay rapidly, with a rate that scales as $1/Re$. Beyond the vorticity diffusion length, $l_v =\sqrt{\eta/(f\rho)} \approx 140$ $\mu$m, which lies outside the field of view of our experiments and simulations, fluid inertia overtakes the viscous effect due to the faster decay of the viscous force with distance \cite{Wei_2019_inertia,Wei_2021_inertia}.  

\begin{figure*}[t]
    \centering
    \includegraphics[width=1\linewidth]{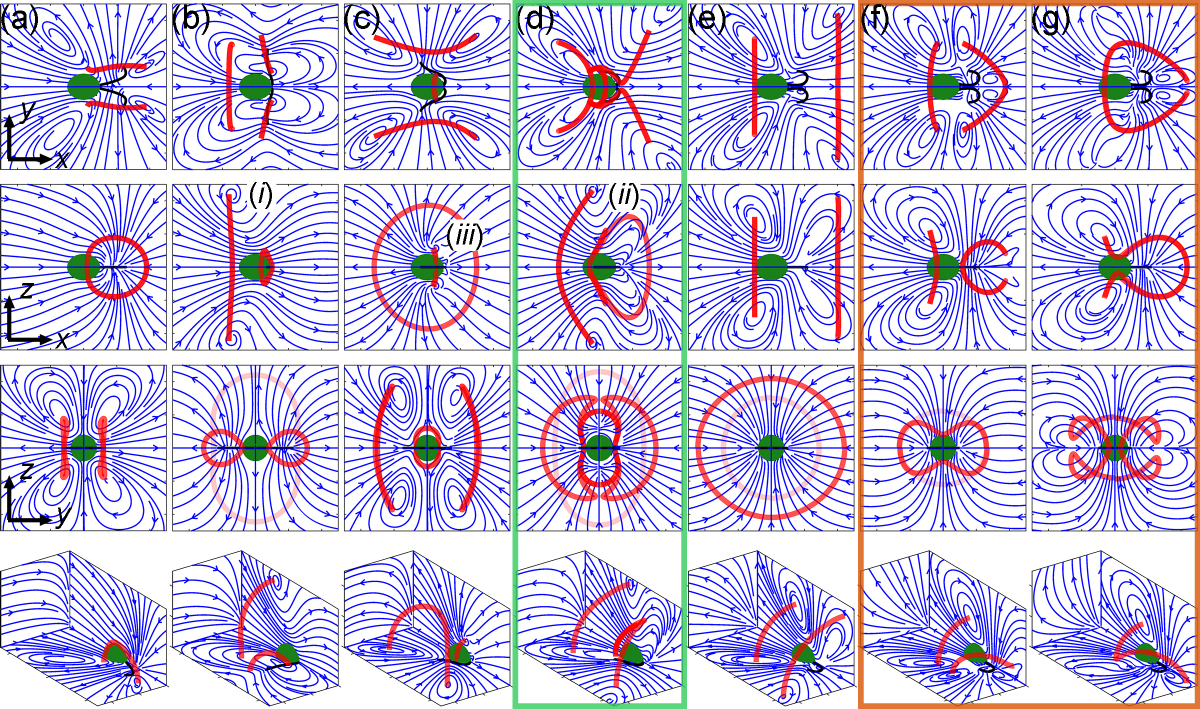}
    \caption{Temporal evolution of 3D vortex structure over one beat cycle from the regularized Stokeslet simulations. Normalized times for (a)-(g) are $t/T = 0$, 0.35, 0.55, 0.6, 0.9, 0.9375, 0.94, where $t=0$ marks the start of the power stroke and $T$ denotes the beating period. Blue lines represent velocity streamlines. Red lines trace vortex lines through the cores of vortices, which approximate the core lines of vortex rings. The top three rows provide orthogonal views of the vortex lines in three different planes, whereas the bottom row shows a 3D perspective of the lines in a quadrant with $y < 0$ and $z > 0$. The third row provides a frontal view, looking directly toward the alga, with lines farther from the viewer rendered in lighter shades. Vortices (i), (ii), and (iii) are labeled at the phases when they first emerge in the $x$–$z$ plane. Transitions between the power and recovery strokes and between the recovery to power strokes are highlighted by the green and orange boxes, respectively. See also Movie S2.}
    \label{fig:figure4}
\end{figure*}

\subsubsection{Vortex dynamics} 

The time-averaged flow reveals a static vortex ring centered around the alga (Figs.~\ref{fig:figure2}(b) and (d)). Micro-vortex rings are also identified in the time-resolved flow, which display complex spatiotemporal patterns (Fig.~\ref{fig:figure4}, Movie S2). At the onset of the power stroke, two vortex rings, one on each side of the alga, are observed (Fig.~\ref{fig:figure4}(a)). As the flagella beat, these rings rotates, completing a half-circle ($\pi$) through the power stroke (Figs.~\ref{fig:figure4}(b) and (c)). A third vortex ring, oriented perpendicular to the $x$-axis, also emerges, first forming behind the algal body and then shrinking as it moves forward during the power stroke. The vortex-ring configuration during the recovery stroke shows less variation: two rings, one in front of and one behind the alga, remain relatively stationary throughout the stroke (Fig.~\ref{fig:figure4}(e)). The most interesting vortex dynamics arise during the transitions from the power to the recovery stroke (Fig.~\ref{fig:figure4}(d)) and from the recovery back to the power stroke (Figs.~\ref{fig:figure4}(f) and (g)). These transitions drive the switch between puller and pusher swimming modes of the alga \cite{Guasto_2010_algalflow,Klindt_2015_simulation}.

\underline{(\textit{i}) Power-to-recovery transition:} As the alga switches from the power to the recovery stroke, the two lateral vortex core lines in Fig.~\ref{fig:figure4}(c) break apart. The front segments of the broken lines reconnect with the middle vortex ring near the body, while the rear segments reconnect with each other forming a new ring (Fig.~\ref{fig:figure4}(d)). The reconnection of the rear segments is synchronized with the emergence of Vortex (ii) in the $x$-$z$ plane. As a result, the three separate vortex rings in Fig.~\ref{fig:figure4}(c) evolve into one isolated vortex ring in the back and a triple torus in the front (Fig.~\ref{fig:figure4}(d)). The cutting and reconnection of vortex rings signify a topological change in the flow structure. Such a change of flow topology has been reported in inertia-driven flows at high $Re$, where cutting and reconnection of vortex lines unlink originally interconnected vortex rings \cite{Ricca_1996_flowtopology,Kleckner_2013_vortex,Yao_2022_vortexringreview}. Although extensively studied, vortex reconnection and the topological change of flow remain an active research topic in fluid mechanics \cite{Yao_2022_vortexringreview}. Few laboratory-scale experimental systems are available to probe the process \cite{Kleckner_2013_vortex}. Here, our observations at zero $Re$ present a new platform for exploring this phenomenon in a  regime where inertia is negligible and topological change is driven solely by moving boundaries.

\underline{(\textit{ii}) Recovery-to-power transition:} The transition from the recovery to power stroke occurs much faster over a short interval of $\sim 0.04T$, triggering a dramatic reorganization of the vortex structure---from a front–back vortex ring pair in Fig.~\ref{fig:figure4}(e) to a left–right vortex ring pair in Fig.~\ref{fig:figure4}(a). The transition proceeds in the following sequence (Movie S3): First, the front vortex ring in Fig.~\ref{fig:figure4}(e) bulges outward and narrows down near the $x$-axis, producing a distorted horizontal figure-eight shape (Fig.~\ref{fig:figure4}(f)). This structure eventually pinches off at the neck into two distinct rings, which then rotate away from each other and subsequently reconnect with the back ring, merging into a single ring with a highly twisted 3D geometry (Fig.~\ref{fig:figure4}(g)). Finally, the rear portion of this folded ring breaks off and disappears, leaving behind two side rings, completing the transformation of the vortex structure from Fig.~\ref{fig:figure4}(e) to (a). Thus, driven by the cutting and reconnection of vortex rings, the recovery-to-power transition is also accompanied by a topological change in the flow.

The swimming mode---whether puller or pusher---is one of the most important characteristics governing the collective dynamics of active fluids and has therefore been extensively studied as a defining feature of self-propelled active particles \cite{Marchetti_2013_review,Saintillan_2018_rheology}. Here, we show that the transitions between these two swimming modes induce not only a reversal of flow direction but also topological changes in the underlying flow structure. It remains an open question how these topological changes modulate inter-algal interactions and influence the collective dynamics and rheology of algal suspensions \cite{Rafai_2010_rheology}.

\begin{figure}[t]
    \includegraphics[width=1\linewidth]{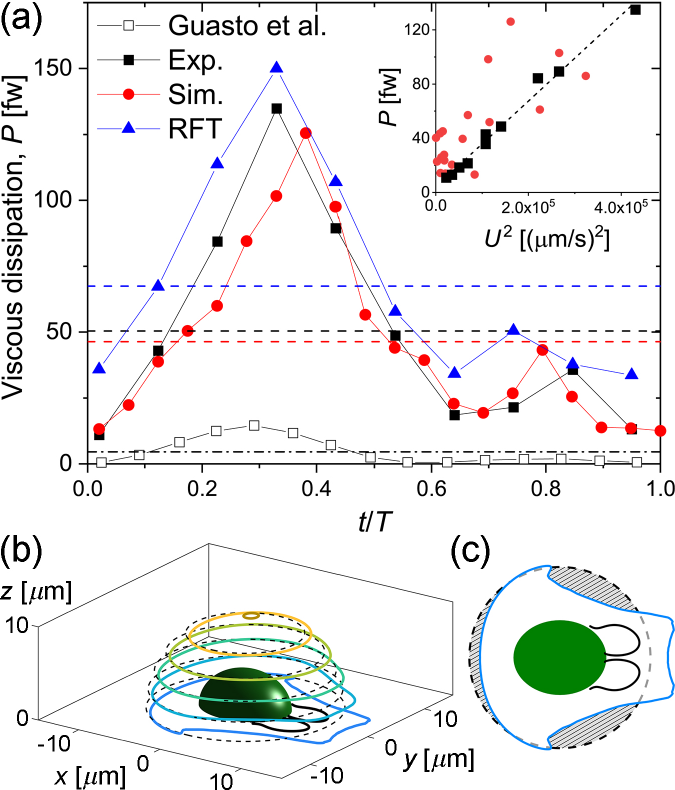}
    \caption{Effect of 3D flow on motility-related algal physiology. (a) Viscous dissipation, $P(t)$, over a flagellar beat cycle. The plot includes experimental and numerical results from this study, along with the estimate provided by Guasto et al. based on a 3D extension of 2D flow field \cite{Guasto_2010_algalflow}. An estimate from resistive force theory (RFT) is also shown. Dashed lines of matching color indicate the cycle-averaged dissipation ($\langle P\rangle = 50.5$ fw from experiments, 46.4 fw from simulations, and 67.5 fw from RFT), while the dash-dotted line shows the cycle-averaged dissipation based on the 2D flow field ($\langle P \rangle = 4.5$ fw). Time $t$ is normalized by the flagellar beating period $T$. Inset: $P(t)$ versus $U(t)^2$, where $U(t)$ is the instantaneous swimming speed. The dashed line is a linear fit to the experimental data. (b) Deformation of a closed material spherical shell induced by flagellar beating. Dashed lines indicate cross-sections of the undeformed shell at various heights $z$ at $t = 0$, while solid lines show the corresponding deformed material lines at $t = T$. From bottom to top, $z = 0, 2, 4, 6, 8$ and $10$ $\mu$m. The cross-section in the flagellar plane ($z = 0$) are shown separately in (c), with the gray region highlighting material points that are drawn closer to the cell after one beat cycle.}
    \label{fig:figure5}
\end{figure}

\section{Biological Implication}

Beyond its relevance to fluid mechanics, the 3D flow has profound biological implications. We examine key physiological metrics of \textit{C. reinhardtii} associated with motility and highlight discrepancies between results derived from our full 3D flow field and those previously reported based on 2D flow measurements.

\subsection{Energy expenditure and swimming efficiency}

Without inertia, an alga expends all the mechanical energy to overcome fluid viscous dissipation. We compute the temporal variation of viscous dissipation over one beat cycle from the time-resolved 3D flow field, $P(t)=\int 2\eta \boldsymbol{\Gamma}:\boldsymbol{\Gamma} d V$, where $\boldsymbol{\Gamma}= \frac{1}{2}[\nabla \boldsymbol{u}+ (\nabla \boldsymbol{u})^T ]$ is the strain rate tensor and $\boldsymbol{u}$ is the flow velocity (Fig.~\ref{fig:figure5}(a)). The cycle-averaged energy dissipation is given by $\langle P \rangle = (1/T)\int_0^T P(t) dt$. Despite the crude approximation underlying RFT, the dissipation estimated based on the theory, $P(t)=2\int_0^l \boldsymbol{f}(s,t)\cdot\boldsymbol{v}(s,t)ds$, shows reasonable agreement with experiments and simulations, where $\boldsymbol{f}(s,t)$ and $\boldsymbol{v}(s,t)$ denote the force density and velocity of a flagellar segment at arclength $s$ along a filament of length $l$ at time $t$. By contrast, the dissipation based on a 3D extension of the 2D flow field, which neglects the 3D vortex dynamics, is more than an order of magnitude smaller \cite{Guasto_2010_algalflow}. Notwithstanding this large discrepancy, the energy expenditure of the alga remains nearly proportional to $U(t)^2$ over a beat cycle (Fig.~\ref{fig:figure5}(a) inset), consistent with both previous findings based on 2D flow \cite{Guasto_2010_algalflow} and the general expectation for energy scaling at low $Re$ \cite{Dusenbery_2009_book}. Here, $U(t)$ is the instantaneous velocity of the alga at time $t$.

From $\langle P\rangle$, we compute the alga's swimming efficiency, 
\begin{equation}
\epsilon_s = \frac{\xi \langle U\rangle^2}{\langle P \rangle},     
\end{equation}
defined as the ratio of the work required to pull the alga body at the average swimming speed $\langle U\rangle$ to the cycle-average mechanical work exerted by the flagella $\langle P \rangle$ \cite{Lauga_2020_book,Tam_2011_optimization}, where the drag coefficient $\xi=6.3\pi \eta c$. The swimming efficiency of the alga based on the 3D flow field, $\epsilon_s = 2.6\%$ ($2.1\%$ from simulations), is slightly higher but still comparable to that of other microorganisms, such as spermatozoa (0.3--1.5\%) \cite{GILLIES_2009_spermefficiency} and flagellated bacteria ($\sim 2\%$) \cite{Kamdar_2023_bacteria}. In contrast, calculations based on the 2D flow field without the 3D vortex structure yield a very high efficiency of 29\%.

\subsection{Feeding efficiency} 
The P{\'e}clet number of the swimming alga is $Pe = 2\langle U \rangle c/D = 1.15$, where $D \approx 2 \times 10^{-5}$ cm$^2$/s is the diffusion coefficient of small molecules (e.g. carbon dioxide or oxygen) in water at room temperature. Thus, advective transport from swimming can substantially alter the nutrient distribution around the alga, thereby affecting its nutrient uptake. Following a method proposed in \cite{Tam_2011_optimization}, we assess the alga's feeding efficiency by quantifying how effectively the flagellar beating draws fluid from farther away toward the cell body. Specifically, we estimate the volume of fluid drawn toward the algal body over one beating cycle, $V$ (Figs.~\ref{fig:figure5}(b) and (c), Appendix E), and define the feeding efficiency $\epsilon_f$ as the ratio of the power required to transport this volume to the mean mechanical power output $\langle P \rangle$: 
\begin{equation}
    \epsilon_f = \frac{5\pi\eta r (rf)^2}{\langle P\rangle},
\end{equation} 
where $r = \sqrt[3]{V}$ is the characteristic length scale of the inbound volume and $rf$ represents the characteristic flow speed.

The feeding efficiency based on the 3D flow field is $\epsilon_f = 14.8\%$ (17.1\% from simulations). For comparison, we also estimate $\epsilon_f$ from the inbound 2D area in the $x$-$y$ plane (Fig.~\ref{fig:figure5}(c)), which yields 13.3\% (14.5\% from simulations), closely matching the optimal feeding efficiency of 13.5\% predicted by a 2D flow analysis \cite{Tam_2011_optimization}. Thus, the 3D flow achieves a feeding efficiency that exceeds that from the 2D analysis---by 11\% in experiments and 18\% in simulations. Lastly, to illustrate the effect of vortex dynamics in nutrient uptake, we further compute $\epsilon_f$ for the static three-Stokeslet model without flagellar beating, whose flow field lacks traveling vortices. In this case, the feeding efficiency drops to $\epsilon_f = 9.7\%$, underscoring the significant role of vortices in boosting nutrient uptake.

\section{Conclusion and Outlook}

We present a direct measurement of the 3D flow field around a freely swimming unicellular microorganism, \textit{C. reinhardtii}. Supported by numerical simulations and hydrodynamic modeling, we quantitatively analyze both the time-averaged and time-resolved flow fields of this model organism. Our results demonstrate how well-established 2D flow features are integrated into and shape the full 3D flow structure of the alga. More importantly, we reveal unexpected flow phenomena at low $Re$, including small vortex rings, moving micro-vortices, and topological changes in the flow structure. These findings expand the repertoire of low-$Re$ flow processes and deepen our understanding of microhydrodynamics and fluid mechanics of cellular motility. Surprisingly, such rich, intriguing, and unusual fluid phenomena have remained hidden within an extensively studied model system to date. 

The detailed 3D flow field also allows us to address biological questions that were previously inaccessible experimentally. In particular, it enables more accurate evaluation of motility-related physiological metrics of the alga. Both the energy expenditure and swimming efficiency derived from our 3D flow field differ by an order of magnitude from previously reported values based on 2D flow data. Moreover, the 3D flow field predicts a higher algal feeding efficiency than the optimal value previously inferred from 2D analyses. 

The 3D flow field revealed in our study opens new avenues for exploring complex interactions between microswimmers and their fluid environment. For example, the flow of a microswimmer governs the dynamics of passive particles during swimmer–particle encounters and underlies the enhanced diffusion of particles in active bath \cite{Leptos_2009_diffusion,Kurtuldu_2011_diffusion,Jeanneret_2016_entrainment,Yang_2016_diffusion,Kurihara_2017_diffusion,Granek_diffusion_2022}. We conduct a preliminary analysis of the trajectories of spherical particles during their encounters with a swimming alga (SM Sec.~3, Fig.~S8) \cite{supplemental}. At small impact parameters, the stagnation point of the 3D flow promotes direct particle entrainment, producing large particle displacements that give rise to heavy tails in the particle displacement distribution \cite{Jeanneret_2016_entrainment}. Despite the complex near-field flow structures and the periodic switching between puller and pusher modes, particle trajectories exhibit looplike paths that reflect the dipolar nature of the average far-field flow, a salient feature discussed in various models of enhanced particle diffusion \cite{Lin_2011_diffusion,Pushkin_2013_diffusion,Kanazawa_2020_diffusion}. We also identify a critical impact parameter at which the net particle displacement reverses direction. The implications of these particle dynamics for long-time diffusion are the subject of ongoing investigations.        

The 3D flow structures also likely influence interactions between an alga and solid surfaces \cite{Durham_2009_algae}, as well as interactions among multiple algae \cite{Pooley_2007_scattering,Muller_2016_scattering}---processes that inherently occur in three dimensions. Moreover, the flow field revealed in our study exhibits complex 3D shear profiles (Figs.~\ref{fig:figure3} and \ref{fig:figure4}) that are expected to induce nontrivial rheological responses in non-Newtonian fluids, thereby affecting microbial locomotion in complex media \cite{Qin_2015_viscoelastics,Li_2017_simulation}. Elucidating how 3D flows mediate cell–wall and cell–cell interactions and modify cellular dynamics in heterogeneous environments, represents an exciting direction for future research.

Our pilot study of the 3D flow field of a freely swimming alga provides only a glimpse into the intricate fluid dynamics occurring at microscopic scales. Although digital in-line holography is a well-established imaging technique, it had not previously been applied to capture the 3D flow fields of swimming microorganisms. Our measurements demonstrate the potential of this technique for measuring the 3D flow fields of free-swimming microorganisms with diverse gaits, supporting its broader use in microbiology and biophysics research.

\textit{Data availability} — The data that support the findings of this article are openly available \cite{data_availability}.

\begin{acknowledgments}
We thank Pete Lefebvre for help with the algae culture, Da Wei for help in data analysis, and Jane Wang for fruitful discussion. The research was supported by David and Lucile Packard Foundation and the U.S. National Science Foundation (NSF), BMMB-2242095 and 2242096/2438345. S.L. was supported by NSF, DMS-1853591 and the Charles Phelps Taft Research Center at the University of Cincinnati. W.L. was supported by the National Institute for Mathematical Sciences Grant funded by the Korean government (B25910000). Y.K. was supported by the National Research Foundation of Korea Grant funded by the Korean government (RS-2023-00247232). Holographic reconstruction was conducted at the Minnesota Supercomputing Institute at the University of Minnesota.
\end{acknowledgments}

\appendix

\section{Algae}
We use wild-type \textit{Chlamydomonas reinhardtii} (CC-125) in our study, which are cultivated axenically in minimum (M1) medium agar plates, a low ionic strength medium that promotes the alga's motility. The culture is maintained in a light-controlled chamber at room temperature, illuminated by a sun lamp (1500 lumens) on a 14-hour light/10-hour dark diurnal cycle. An algal suspension is prepared by transferring alga from the agar plate to 1 mL of liquid M1 medium. The liquid culture is allowed to grow for 48 hours in the light-controlled chamber. At the end of this period, the suspension is diluted to an algal concentration about 2000 cells/mL and left to rest for 2-3 hours for the culture to complete its logarithmic growth phase. The final suspension is then mixed with a suspension of 1-$\mu$m-diameter spherical polystyrene (PS) particles (0.02\% v/v), which serve as tracers for flow visualization.

\section{Holographic microscopy}
We achieve 3D flow visualization around a freely swimming alga, utilizing a custom-built digital inline holographic microscopy (DIHM) setup (SM Fig.~S4) \cite{supplemental}. We use an inverted optical microscope (Nikon Ti-E) equipped with a high-speed camera (iX Camera, i-Speed 220) and a $40\times$ water-immersion lens. A blue fiber-optic laser ($\lambda = 452$ nm, QPhotonics QFLD-450-10S) is used as our coherent illumination source, which is collimated by a pinhole aperture and a reversed $20\times$ objective lens. A hologram is generated when the collimated laser beam interferes with the light scattered from the tracer particles. Videos of holograms are captured at 500 frames per second (FPS), enabling us to resolve 10 flagellar phases within one beat cycle of flagella. Our analysis focuses exclusively on holograms when both flagella beat synchronously and remain in focus.

The experimental chamber is constructed using a microscope cover slide and three cover slips, all of which are base-washed to minimize particle and algae adhesion. The chamber dimensions are 170 $\mu$m in height, 2 cm in length, and 0.5 cm in width. The holography focal plane is centered within the chamber, parallel to but offset from the top and bottom walls of the chamber. To prevent drift caused by ambient airflow, the chamber is sealed with UV-curable adhesive after being filled with samples. Algae stay active in the sealed chamber for $\sim 2$ hours, during which we conduct our experiments. 

We reconstruct the recorded holograms to obtain 3D particle positions at different times. Conceptually, reconstruction involves illuminating a hologram with light of the same wavelength as the original illumination source (a reconstruction wave). This process can be executed computationally by applying a diffraction equation that convolves the reconstruction wave with the raw hologram. Specifically, we use a custom MATLAB algorithm based on the Rayleigh-Sommerfeld diffraction equation \cite{Toloui_2015_holography}. The convolution generates a series of 2D darkfield images representing light intensity at different $z$ positions, which together form a 3D intensity field containing the 3D information of particles. Using the algorithm, we reconstruct the intensity field and obtain particle trajectories up to 30 $\mu$m away from the focal plane.

A particle appears as a blob of high intensity pixels in the 3D intensity field. After applying a threshold to reduce noise, the 3D position of each particle is determined by calculating the centroid of the blob. We use the open-source Python package, \textit{trackpy}, to connect particle positions across different frames, creating 3D particle trajectories. A centered finite-difference scheme is then used to calculate the instantaneous velocity of particles at different times in the lab frame. 

To calibrate and assess the spatial resolution of our method, we measure the diffusion of PS tracers in water without algae. The mean-squared displacements (MSDs) of the particles exhibit 3D Brownian diffusion (SM Fig.~S5) \cite{supplemental}. By fitting the MSDs in each direction with $\langle x_i^2\rangle= 2Dt+2\Delta_\epsilon^2$, we estimate the errors $\Delta_\epsilon$ to be 40 nm in the $x$- and $y$-directions and 55 nm in the $z$-direction \cite{Crocker_1996_PTV}.

\section{3D flow field}

\subsection{Reference frame transformation}

To obtain the flow field around a freely swimming alga, we convert the flow field from the lab frame to the alga-centered frame. In this frame, the alga is located at the origin and oriented along the $+x$ direction. The locations and orientations of the velocity vectors are transformed accordingly, while the magnitudes of the velocity vectors remain the same as in the lab frame. Since holograms are captured only when the alga is in the focal plane, we begin by tracking the alga’s position $\boldsymbol{x}_{b,l} =(x_{b,l},y_{b,l})$ within this plane from the raw holograms. The swimming direction of the alga with respect to the $+x$ direction, $\theta_{b,l}$, is determined by fitting the trajectory with a linear function locally over the time interval of one beating cycle.
We then transform a particle trajectory from the lab frame to the alga-centered frame:
\begin{equation} \label{eq:referenceframe}
 \boldsymbol{x}_{p,a} = \boldsymbol{A}\cdot\left(\boldsymbol{x}_{p,l} - \boldsymbol{x}_{b,l}\right),
\end{equation}
where $\boldsymbol{x}_{p,l} = (x_{p,l},y_{p,l})$ is the position of a particle in the lab frame, $\boldsymbol{x}_{p,a}=(x_{p,a},y_{p,a})$ is the particle position in the alga-centered frame, and the transformation matrix
\begin{equation}
    \boldsymbol{A} = \begin{bmatrix}
    \cos \theta_{b,l} & -\sin \theta_{b,l} \\ \sin \theta_{b,l} & \cos \theta_{b,l}
\end{bmatrix}.
\end{equation}
Using Eq.~\ref{eq:referenceframe}, the location of the velocity vector, $\boldsymbol{u}_{p,l}=(u_{p,l},v_{p,l},w_{p,l})$, obtained from the lab-frame trajectory is mapped onto the corresponding position in the alga-centered frame. The vector $\boldsymbol{u}_{p,l}$ is finally rotated to align the swimming direction of the alga in the $+x$ direction:
\begin{equation}       \boldsymbol{u}_{p,a}=\boldsymbol{B}\cdot\boldsymbol{u}_{p,l},
\end{equation}
which gives the velocity vector of the tracer particle in the alga-centered frame, $\boldsymbol{u}_{p,a}=(u_{p,a},v_{p,a},w_{p,a})$ with the matrix
\begin{equation}
    \boldsymbol{B} = \begin{bmatrix}
        \cos \theta_{b,l} & -\sin \theta_{b,l} & 0 \\ \sin \theta_{b,l} & \cos \theta_{b,l} & 0 \\ 0 & 0 & 1
    \end{bmatrix}.
\end{equation}

To construct the 3D flow field, velocity vectors in the alga-centered frame are binned into 1-$\mu$m-cube voxel grid. Averaging all the vectors within a voxel yields the velocity of the flow in the position of the voxel. To increase the spatial resolution and smooth the flow field, each voxel has an overlap of 0.25 $\mu$m with its neighboring voxels. 

\subsection{Challenges and solutions} 

The difficulties of tracking the 2D flow field around a fast, irregularly moving microorganism with micron-scale spatial resolution and millisecond temporal precision are well summarized in Ref.~\cite{Goldstein_2015_review}. Extending such measurements to 3D poses a significantly greater challenge beyond the already demanding 2D measurements. Although a large data set is required in both 2D and 3D flow measurements to overcome the stochastic nature of microorganism swimming and minimize thermal noise of small tracers \cite{Goldstein_2015_review}, the size of data required for the 3D flow field is substantially bigger. To begin with, holography limits the maximum allowable volume fraction of tracers, which is lower than that typically used in 2D particle tracking velocimetry. A high tracer concentration results in strong speckle noise, which degrades the quality of holographic reconstruction. Additionally, while 2D projection allows tracer trajectories from different heights to be superimposed on a single pixel in the 2D flow field, the 3D flow field requires a large number of trajectories counted independently within each voxel. To address these challenges, we utilize the symmetry and periodicity of the algal flow to improve the statistics and signal-to-noise ratio of our measurements.

First, the synchronous beating of flagella dictates the symmetry of the flow field about the central axis of the alga. Moreover, we neglect the weak secondary flow due to the out-of-plane beating of flagella \cite{Cortese_2021_3Dflagella} and focus on the primary flow induced by their dominant in-plane beating. Under this approximation, the 3D flow field possesses an additional mirror symmetry with respect to the $x$-$y$ plane. The full $C_{2v}$ symmetry of the flow allows us to average the flow field across four quadrants, effectively quadrupling the data volume. To verify the approximation, we conduct immersed boundary simulations on an alga performing out-of-plane flagellar beating. The resulting flow field is quantitatively similar to that of an alga with only in-plane flagellar beating, retaining all the key flow features reported here (SM Sec.~2C) \cite{supplemental}. 

Second, as an alga beats its flagella at approximately 50 Hz, the motions of tracer particles driven by the algal flow should also oscillate at the same frequency. To improve the signal-to-noise ratio and minimize the influence of random thermal noise, we apply a band-pass filter to our velocity data \cite{Wei_2019_inertia}. To account for variations in the beat frequency among different algae, we first apply a band-pass filter with a range of 40 to 60 Hz to each of the three velocity components. We then identify the dominant frequency by performing a Fourier transform on the filtered velocity data. Next, we apply a narrower band-pass filter centered on the dominant frequency, with a range of $\pm 2.5$ Hz. If the dominant frequencies differ among the three velocity components, we set the band at $\pm 2.5$ Hz around the mean of the dominant frequencies. The procedure filters out abnormal high velocities due to Brownian motion and significantly improves the signal-to-noise ratio across all trajectories, especially those more than 20 $\mu$m from the alga.

Finally, we analyze over 2,000 flagellar beat cycles and more than 50,000 tracer trajectories to obtain the robust 3D flow field. As a result, each voxel in our 3D flow field contains on average more than 45 velocity vectors. When constructing the time-averaged flow, we average all the vectors within a voxel regardless the flagellar beat phase. For the time-resolved flow field, we sort tracer particle trajectories by flagellar phases. Specifically, we fit the displacement of the algal body, $x$, as a function of time $t$ in the form of $x(t)=A\exp(Bt)\sin(Ct)+Dt+E$. The fitted trajectory over a single cycle is then divided into ten discrete time intervals, with each interval representing a distinct flagellar phase (SM Fig.~S7) \cite{supplemental}. 

\section{Identification of vortex core} 
Time-resolved flow fields from experiments and simulations reveal the formation and propagation of vortices in three orthogonal planes. To connect vortices across different planes and analyze their intrinsic 3D structure and dynamics, we identify 3D vortices and their core lines from the flow field. Specifically, we apply a method originally developed for turbulent flow, which is also applicable for flows at low $Re$ \cite{Jeong_1995_vortexidentify}. The method identifies the core of a 3D vortex by tracking local pressure minima due to vortical motion in 2D planes. These pressure minima can be determined from the eigenvalues of the tensor, $\boldsymbol{\Lambda}= \boldsymbol{\Gamma}^2+ \boldsymbol{\Omega}^2$, where $\boldsymbol{\Gamma} = \frac{1}{2}[\nabla \boldsymbol{u}+(\nabla \boldsymbol{u})^T]$ and $\boldsymbol{\Omega}= \frac{1}{2}[\nabla \boldsymbol{u}-(\nabla \boldsymbol{u})^T ]$ are the symmetric and antisymmetric components of the velocity gradient, respectively. As $\boldsymbol{\Lambda}$ is symmetric, it has three real eigenvalues $(\lambda_1,\lambda_2,\lambda_3)$ arranged in descending order, $\lambda_1 \geq \lambda_2 \geq \lambda_3$. A local pressure minimum induced by vortical motion in an eigenplane corresponds to two negative eigenvalues of $\boldsymbol{\Lambda}$. Thus, the boundary of the vortex core is associated with the isosurface of $\lambda_2=0$. Notably, we find that the $\lambda_2=0$ isosurface aligns well with an isosurface of swirling strength \cite{Zhou_1999_hairpin}, further validating the approach. SM Figure~S3 shows a representative example of the $\lambda_2=0$ isosurface from the flow field at the start of the power stroke corresponding to the phase in Fig.~\ref{fig:figure4}(a) \cite{supplemental}. A vortex ring can be clearly identified from the topology of the isosurface. 

To further pinpoint vortex core lines when vortex rings are present, we implement two additional criteria in the region contained within the isosurface: local minimum of velocity magnitude $|\boldsymbol{u}|$, and local maximum of relative vorticity $|\boldsymbol{\omega}|/|\boldsymbol{u}|$. We scan 2D slices of the flow field along all three orthogonal planes, identifying points that meet these extrema conditions within the $\lambda_2=0$ isosurface. Finally, we search for points on the $x$-$y$ and $x$-$z$ planes that are visually confirmed to be at the center of vortices. Using them as the starting points, we plot vortex lines, i.e., streamlines of the vorticity vector field, that approximate the locations of vortex cores during phases with prominent vortex rings and trace the centers of local vortices in other phases (Fig.~\ref{fig:figure4}).

\section{Feeding efficiency}

To estimate the feeding efficiency of the alga, we follow the method proposed in \cite{Tam_2011_optimization} and extend it from 2D to 3D. Specifically, we measure the volume of fluid $V$ that is displaced towards the alga over one cycle of flagellar beating. Consider a closed material surface $r_0(\theta,\phi,t)$, where $\theta$ and $\phi$ are the polar and azimuthal angles. At $t = 0$, this surface forms a spherical shell of radius $r_0(\theta,\phi,0)= R_0$ around the alga. We set $R_0 = 2c = 10$ $\mu$m, following \cite{Tam_2011_optimization}. Tracking the movement of material points over one beat cycle, we obtain the deformed surface $r_0(\theta,\phi,T)$ (Figs.~\ref{fig:figure5}(b) and (c)). The volume of the region defined by points where $r_0(\theta,\phi,T) < r < r_0(\theta,\phi,0)$ is denoted as $V$. This volume represents material points drawn closer to the cell surface during the stroke, providing a measure of the inward volumetric flow rate. 

\section{Hydrodynamic modeling and simulations}

Details of our hydrodynamic models---the three-Stokeslet model (Figs.~\ref{fig:figure2}(c) and (d)), the three-Oscillet model (Movie~S1), and the dynamic three-Stokeslet model (Fig.~\ref{fig:figure3}(b))---are provided in the Supplementary Material (Sec.~1) \cite{supplemental}. Descriptions of our numerical simulations, including the regularized Stokeslet method (Figs.~\ref{fig:figure3}(c) and \ref{fig:figure4}) and the immersed boundary method, are included in Sec.~2 of the Supplementary Material \cite{supplemental}.


\bibliography{apssamp}

\end{document}